\documentclass[a4paper]{article}
\usepackage{INTERSPEECH2020}
\usepackage{mathrsfs}
\usepackage{subfigure}
\usepackage{graphics}
\usepackage{multirow}
\usepackage{multicol}
\usepackage{amsmath}
\usepackage{booktabs}       
\usepackage{amsfonts}       
\usepackage{nicefrac}       
\usepackage{microtype}
\usepackage{float}
\usepackage{xcolor}
\usepackage{url}
\usepackage[symbol]{footmisc}

\title{Opencpop: A High-Quality Open Source Chinese Popular Song Corpus for Singing Voice Synthesis}
\name{Yu Wang$^{1,2 \ast} $\footnotemark[1], Xinsheng Wang$^{3,4\ast}$\footnotemark[1], Pengcheng Zhu$^{3,4 \dagger} $\footnotemark[2], Jie Wu$^{1 \dagger}$, Hanzhao Li$^{3}$, Heyang Xue$^{3,4}$, Yongmao~Zhang$^{3}$, Lei Xie$^{3}$\footnotemark[2], Mengxiao Bi$^{4}$ \footnotetext[1]{These authors contributed equally to this work.} }
\address{
  $^1$ College of Design and Innovation, Tongji University, China \\
  $^2$ School of Pop Music \& Dance, Shanghai Institute of Visual Arts, China\\
  $^3$Audio, Speech and Language Processing Group (ASLP@NPU), School of Computer Science, Northwestern Polytechnical University, Xi’an, China \\
  $^4$Fuxi AI Lab, NetEase Inc., Hangzhou, China
  }

\email{asterwangyu@126.com, w.xinshawn@gmail.com, zhupengcheng@corp.netease.com, lihanzhao.mail@gmail.com, xueheyang@corp.netease.com, zym@mail.nwpu.edu.cn, wujie@tongji.edu.cn, lxie@nwpu.edu.cn, bimengxiao@corp.netease.com}

\begin{document}

\maketitle

\renewcommand{\thefootnote}{\fnsymbol{footnote}} 
\footnotetext[1]{These authors contributed equally to this work.}
\footnotetext[2]{Corresponding authors.}
\renewcommand{\thefootnote}{\arabic{footnote}}
\begin{abstract}
This paper introduces Opencpop, a publicly available high-quality Mandarin singing corpus designed for singing voice synthesis (SVS). The corpus consists of 100 popular Mandarin songs performed by a female professional singer. Audio files are recorded with studio quality at a sampling rate of 44,100 Hz and the corresponding lyrics and musical scores are provided. All singing recordings have been phonetically annotated with phoneme boundaries and syllable (note) boundaries. To demonstrate the reliability of the released data and to provide a baseline for future research, we built baseline deep neural network-based SVS models and evaluated them with both objective metrics and subjective mean opinion score (MOS) measure. Experimental results show that the best SVS model trained on our database achieves 3.70 MOS, indicating the reliability of the provided corpus. Opencpop is released to the open-source community WeNet\footnote{\url{https://github.com/wenet-e2e/}}, and the corpus, as well as synthesized demos, can be found on the project homepage\footnote{\url{https://wenet.org.cn/opencpop/}}.

\end{abstract}


\section{Introduction}

Singing voice synthesis (SVS), which aims to synthesize singing voices from text and musical score information (e.g., note and tempo), has great potential for many creative applications, e.g., virtual avatars and artistic creation. To promote the development of Mandarin SVS, a high-quality Mandarin SAS corpus, referred to as Opencpop, with fine annotated textual information and musical scores, is released in this paper. 

The development of neural end-to-end text-to-speech (TTS) models~\cite{wang2017tacotron, shen2017natural, li2019neural, sotelo2017char2wav, ren2020fastspeech2, yu2019durian} has greatly promoted speech synthesis. Generally, with a well-trained neural acoustic model~\cite{shen2017natural, ren2020fastspeech2, yu2019durian,miao2021efficienttts} and a neural vocoder~\cite{oord2016wavenet,kim2018flowavenet,prenger2019waveglow,kong2020hifi}, or alternatively using fully end-to-end models~\cite{donahue2020end,weiss2021wave,kim2021conditional} which directly construct wave signals from text input, it is able to synthesize high-quality neutral speech. Recently, much attention has been attracted to synthesizing expressive speech, such as stylized speech~\cite{skerry2018towards,wang2018style}, emotional speech~\cite{lee2017emotional,rabiee2019adjusting, zhu2019controlling,cai2020emotion,um2020emotional,li2021controllable}, and also singing voice~\cite{chen2020hifisinger,liu2021efficientsing}. 

Apart from technological innovating, a proper database is crucial for evaluation and comparison. The end-to-end TTS technology allows to train a model only with audio files and paired textual transcriptions, such as corpora LJ-Speech~\cite{ljspeech17}, VCTK~\cite{veaux2016superseded}, Aishell-3~\cite{shi2020aishell}, and DiDiSpeech~\cite{guo2021didispeech}. However, the case in the SVS task is different, in which the musical score is generally needed. While several open-source SVS databases designed for English~\cite{duan2013nus,sharma2021nhss} or other languages~\cite{tamaru2020jvs} have been released and also lots of research has been conducted towards Mandarin singing voice synthesis~\cite{liu2021efficientsing,lu2020xiaoicesing,huang2021multi}, no accessible public Mandarin SVS corpus with high quality exists, thereby limiting the development of Mandarin SVS. Note that although Chinese singing voice databases OpenSinger~\cite{huang2021multi} and PopCS~\cite{liu2021diffsinger} were released recently, accurate musical score information was missing in both corpora, with which an automatic fundamental frequency detection algorithm is necessary to calculate the rough pitch information. These corpora, thus, are hard to satisfy an SVS system with high quality.

A major cause of this data shortage is the challenge of annotating a high-quality SVS corpus. Different from the corpora designed for speech synthesis, in which only the recordings and corresponding transcriptions are necessary, extra musical score information, such as note and note boundaries, is also necessary for training an SVS model. While the song is generally sung based on the existed musical score, the singing voice is hard to be exactly aligned with the musical score even for a professional singer. Thereby, manual post-processing, such as sound shaping or musical score relabeling, is necessary, which depends on heavy labor from professionals, making it much harder to build such a corpus.

In this paper, we introduce the accessible public Opencpop, which is a high-quality SVS corpus with the phonetically manual-annotated musical score. In this corpus, 100 Mandarin Pop songs are recorded by a professional female singer. All audio is recorded in a professional recording studio with a sampling rate of 44,100 Hz. The total duration of the recording is around 5.2 hours, and the distributions of phoneme, syllable, and pitch will be introduced in the paper, respectively. Besides, based on the new corpus, baseline SVS models are built to test the reliability of the collected database, and also provide the baseline performance for future research. 

The rest of this paper is organized as follows: Section \ref{sc:data} details the corpus creation pipeline from collecting songs to the final audio files. The corpus's overall statistics, including both audio and textual transcripts, are also presented in this section. We introduce the details of training different models based on Opencpop as well as the evaluation and the results in Section \ref{sc:experiments}. Finally, we conclude the paper in Section \ref{sc:conclusions}.

\section{The creation of Opencpop}
\label{sc:data}
\begin{figure}[tbp]
\setlength{\belowcaptionskip}{-0.3cm}
    \centering
    \includegraphics[width=\linewidth]{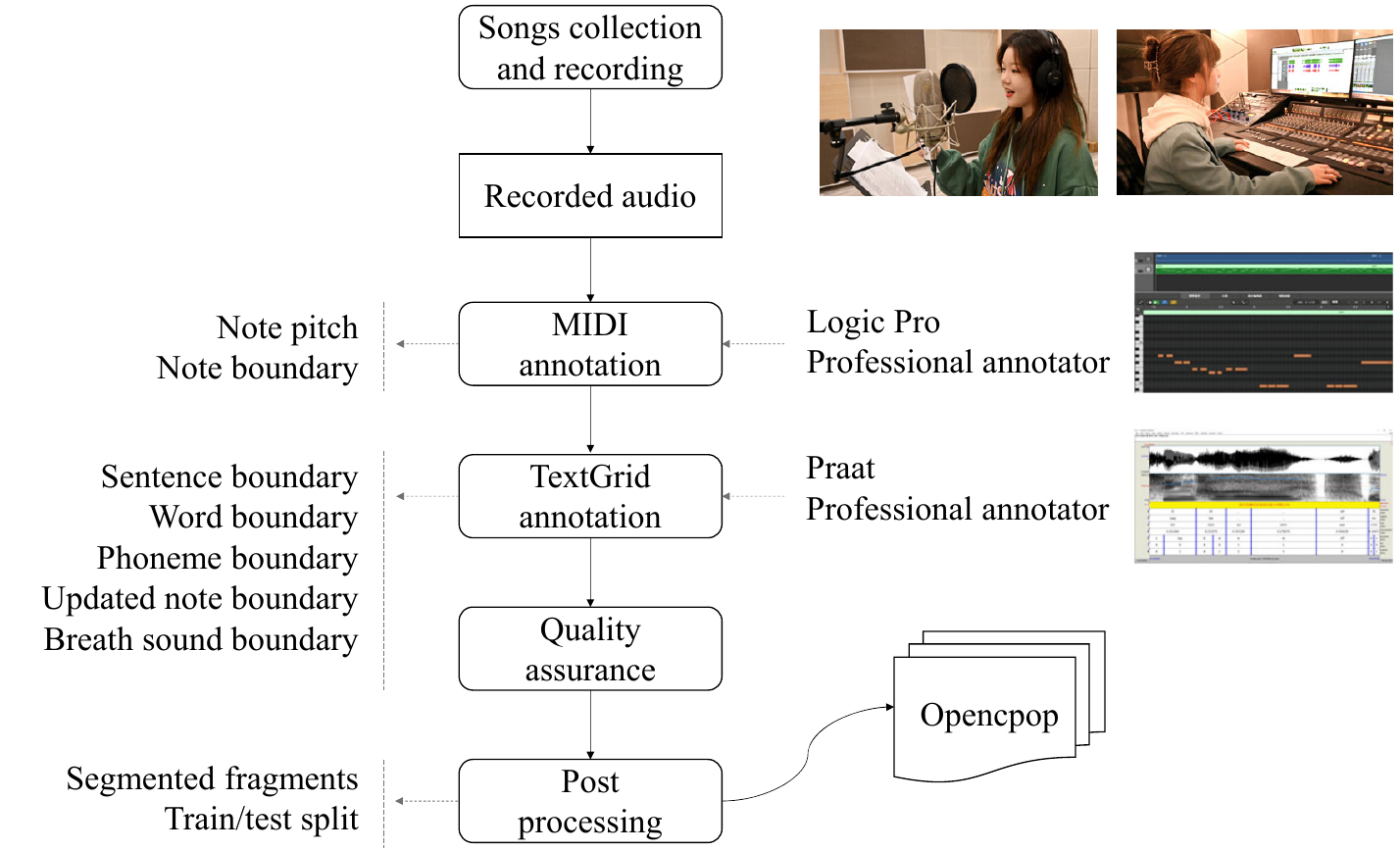}
    \caption{Flowchart for creating the Opencpop database. For clear presentation, here, quality assurance is only illustrated after the TextGrid annotation. In practice, the human double-check exists in every process.}
    \label{fig:flowchart}
\end{figure}

In this section, we will introduce Opencpop, a publicly available Mandarin singing corpus designed for the SVS task, which has fine manually annotated textual information and musical scores. The flowchart of Opencpop creation is shown in Fig.~\ref{fig:flowchart}, and the details of each step for creating this database will be described in this section.
\subsection{Songs and the singer}
\label{sc:song_selection}
All songs were selected according to several music charts for China Pop songs. The originally collected songbook consisted of 300 songs, based on which the final 100 songs were selected with the following processes: 1) any song would be removed from the songbook if its lyric contains non-Chinese characters; 2) the remained songs' beats per minute (BPM) were counted, and songs with the BPM that occurs in relatively lower frequency were chosen to the final song list preferentially; 3) similar to the previous BPM-based rule, remained songs with infrequent phonemes were selected to the final list; 4) the final selected song list was completed by choosing songs that contain as many phonemes as possible. After the above steps, 100 Mandarin Pop songs with sufficient phoneme and BPM coverage were chosen for the recording.

The singer who would perform these songs is a professional young female singer. Her vocal range can perfectly perform all the selected songs. 

\begin{figure}[tbp]
\setlength{\belowcaptionskip}{-0.3cm}
    \centering
    \includegraphics[width=\linewidth]{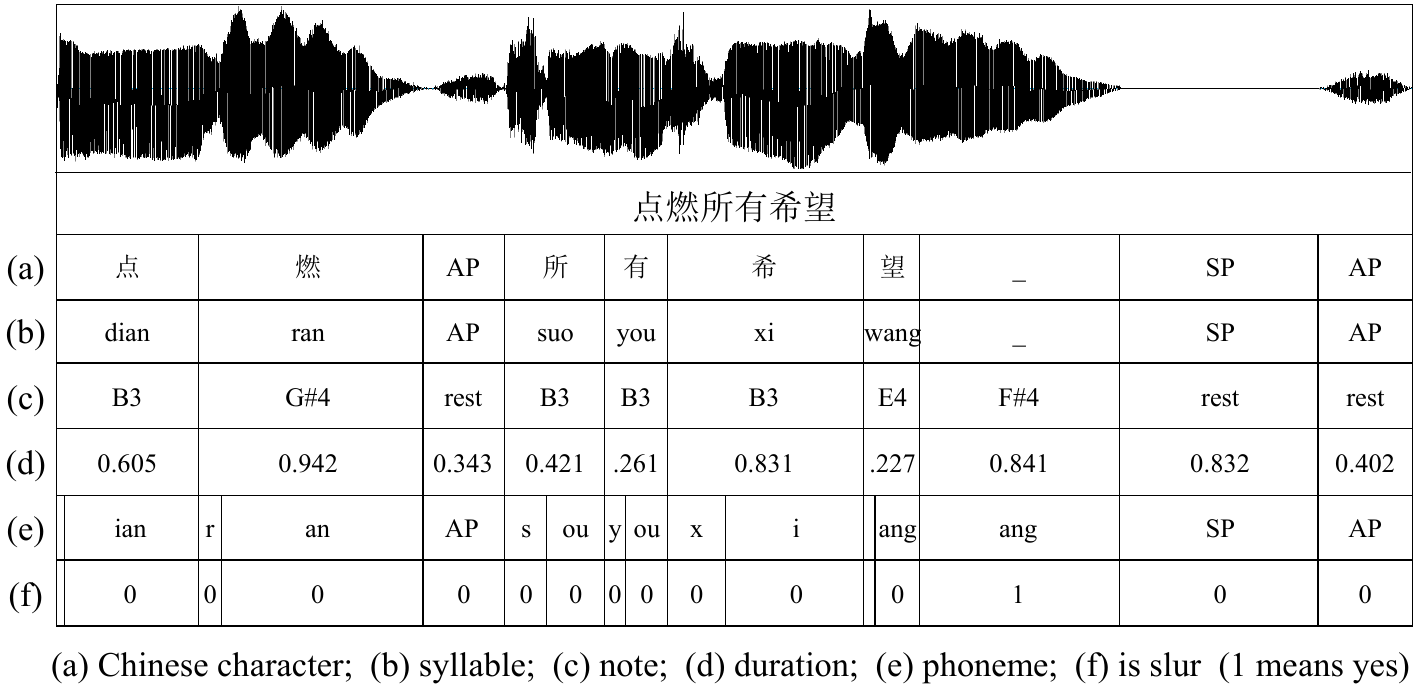}
    \caption{Illustration of a labeled sample. The phoneme ``d" of syllable ``dian" and the phoneme ``w" of syllable ``wang" are omitted due to the limited space. The ``rest" in the note track corresponds to the silence (SP) or aspirate (AP).}
    \label{fig:format}
\end{figure}

\subsection{Recording}

The recording was conducted in a professional studio with sufficiently little reverb, and the singing voice was recorded as an individual track, which means each recorded audio contains pure vocal only, without any background music. During the recording, the singer was equipped with a headphone, which provided the accompanying audio, to ensure the steadiness and accuracy of pitch. All the data were recorded in 44,100 Hz sampling rate with 24 bits per sample in wav format.


\subsection{Labeling}

In the Opencpop database, manually labeled information includes note pitch, note boundary, phoneme boundary, syllable boundary, word boundary, utterance boundary, and the indicator to indicate whether a note is a slur or not. An illustration of a labeled sample is shown in Fig.~\ref{fig:format}. Details to obtain those annotations will be described in this part.

\subsubsection{MIDI annotation}
The musical score, as a written form of a musical composition, normally includes the note pitch, note duration, the tempo, the key signature, etc. The music score that the singer accords to during the recording can be easily transcribed to MIDI format. However, even though our singer is professional in singing, it is non-trivial to ensure the sing voice aligned with the musical score perfectly. Therefore, this music score-based method is hard to create adequate results. Instead, a semi-automatic method was adopted to create the MIDI score based on the recorded audio rather than the original music score. 

To be specific, with the input of audio, Logic Pro\footnote{https://www.apple.com/logic-pro/}, which is a digital audio workstation and MIDI sequencer software, was utilized to automatically create the preliminary musical score. Then this preliminary musical score was manually adjusted by professional annotators to obtain the more accurate but not final musical score. The musical score obtained in this stage would be further tuned during the TextGrid annotation process, which will be introduced in the next subsection.

\subsubsection{TextGrid annotation}
The TextGrid annotation is to label all information illustrated in Fig.~\ref{fig:format}, in which process the note-related annotations, i.e., note pitch and note duration, were taken from the MIDI annotation results, and the note duration (boundary) was fine-tuned to align with the newly labeled syllable boundary. To efficiently label boundaries of pronunciation in various levels, a preliminary alignment between phonemes and the corresponding audio was obtained with Montreal Forced Aligner\footnote{https://github.com/MontrealCorpusTools/Montreal-Forced-Aligner} for automatic songs' phoneme annotation. The achieved rough phoneme boundaries are very friendly for annotators to annotate the final accurate phoneme boundaries in Praat~\cite{boersma2001praat}. Then, the syllable boundaries can be easily obtained based on the phoneme boundary. Here, we argue that the boundary of each syllable is aligned with the boundary of a pitch, which means the pitch boundary achieved in the MIDI annotation process would be adjusted to align with the syllable boundary. 

A case to show different levels' boundaries is presented in Fig.~\ref{fig:format}. As can be seen, different from TTS-oriented textual phoneme sequences that can be directly converted from sentences, in Opencpop database, some phonemes could be repeated due to the slur in the song, for instance, the phoneme ``ang" of syllable ``wang" is repeated one time. The indicator of the slur was also annotated in Opencpop database. Besides, the aspirate was also labeled.

\subsection{Post processing}

\subsubsection{Audio segmentation}
After the annotation, we segmented the recording into smaller fragments for the training of the SVS system. All song audio was first segmented into sentence-level according to the transcription and the annotated boundary information. Then, to avoid the existence of too-long utterances in the final database, any utterances longer than 8 seconds would be further segmented into shorter utterances. During this segmentation process, the end boundary of a \textit{rest} within the utterance (\textit{SP} in Fig.~\ref{fig:format}) would be the priority segmentation point. If no \textit{SP} exists within an utterance that is longer than 8 seconds, the prosodic boundary achieved by prosody analysis would be set as the segmentation point. After dropping silence utterances, we obtained 3,756 utterances finally with a total duration of around 5.2 hours.  

\subsubsection{Training set split}
\begin{figure}[t]
    \centering
    \setlength{\abovecaptionskip}{0.2cm}
    \setlength{\belowcaptionskip}{-0.5cm}
    \includegraphics[width=\linewidth]{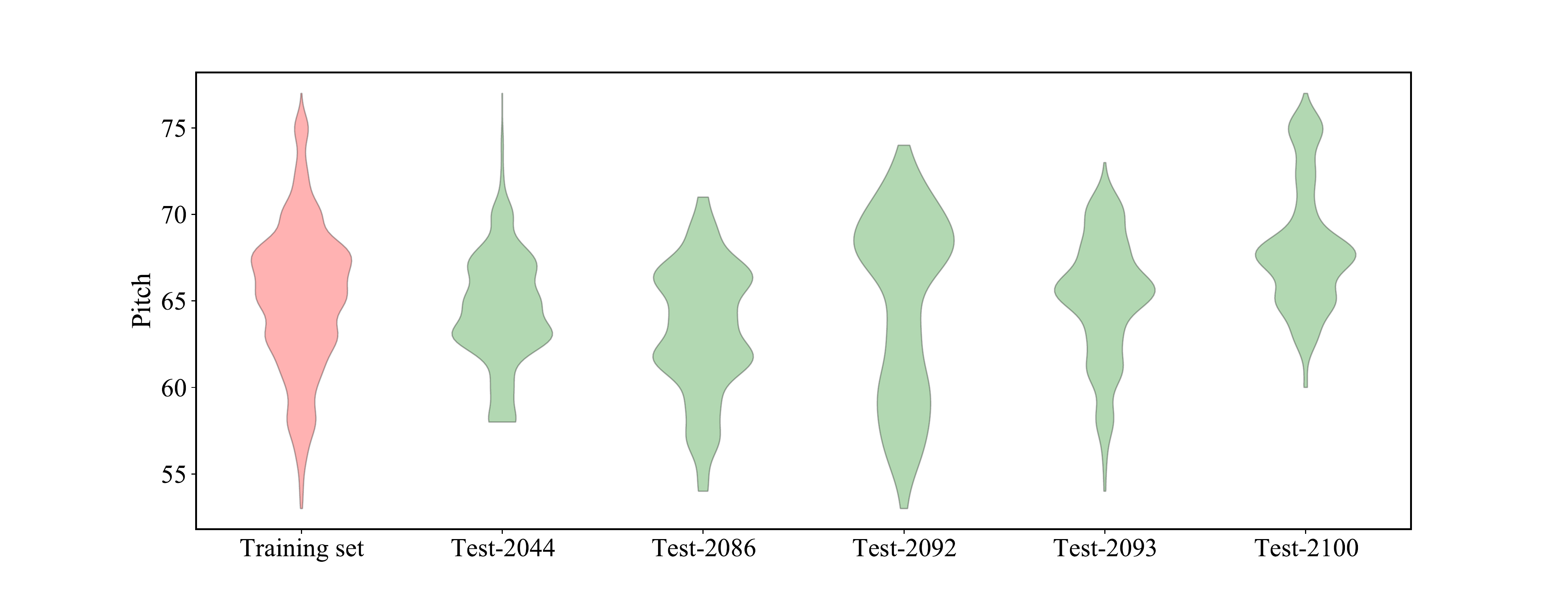}
    \caption{The violin plots of pitch numbers in different subsets. Pitch is presented as MIDI note number, where A4=69=440 Hz.}
    \label{fig:violin}
\end{figure}

For the training and evaluation of an SVS system, 5 songs out of the total 100 songs were handpicked as the test set, which contains 206 utterances, and the remaining 95 songs with 3,550 utterances were set as the training set. The violin plots of pitch numbers in different test songs and the training set are compared in Fig.~\ref{fig:violin}. As can be seen, the selected 5 test songs cover several typical cases, i.e., with relatively high (Test-2100), low (Test-2086 and Test-2092), and medium pitch values (Test-2100), and thus can support a comprehensive evaluation for an SVS system. Details of the Opencpop database's statistics will be introduced in Seciton~\ref{sc:statistics}.

\begin{figure}[t]
    \centering
    \setlength{\abovecaptionskip}{0.1cm}
    \setlength{\belowcaptionskip}{-0.1cm}
    \includegraphics[width=\linewidth]{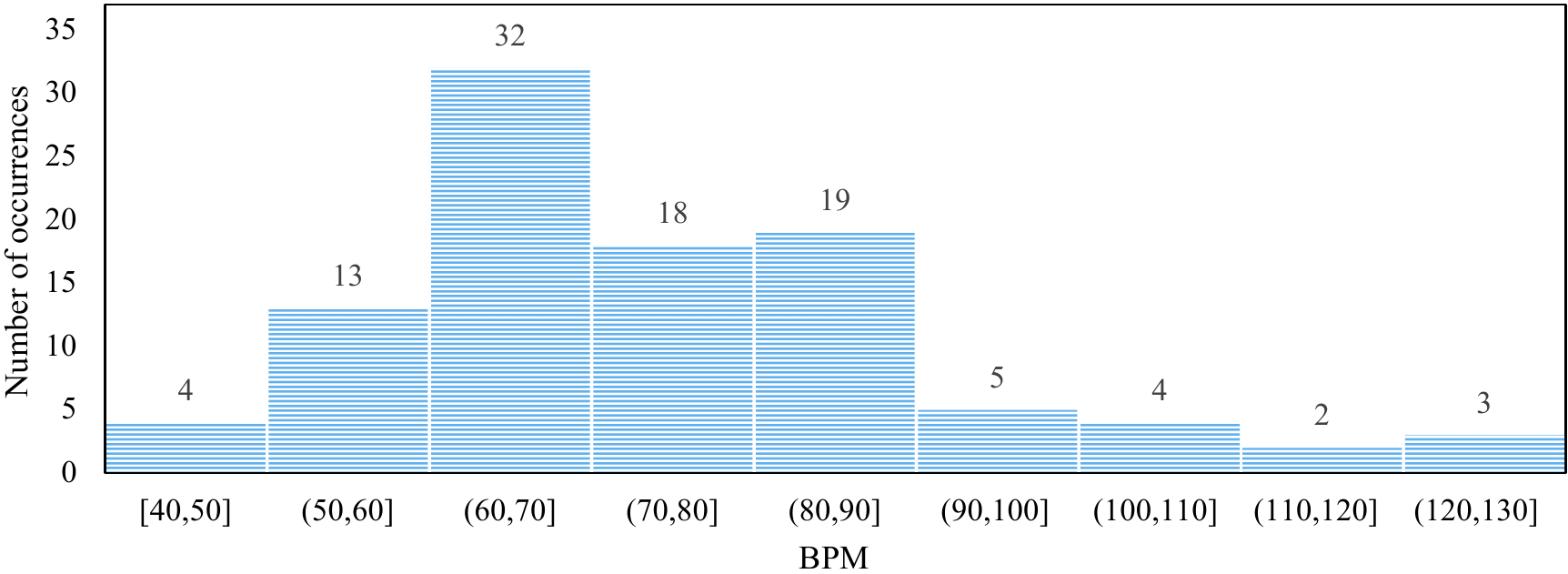}
    \caption{The statistical distribution of BPM.}
    \label{fig:bpm}
\end{figure}

\begin{figure}[t]
    \centering
    \setlength{\abovecaptionskip}{0.1cm}
    \setlength{\belowcaptionskip}{-0.1cm}
    \includegraphics[width=\linewidth]{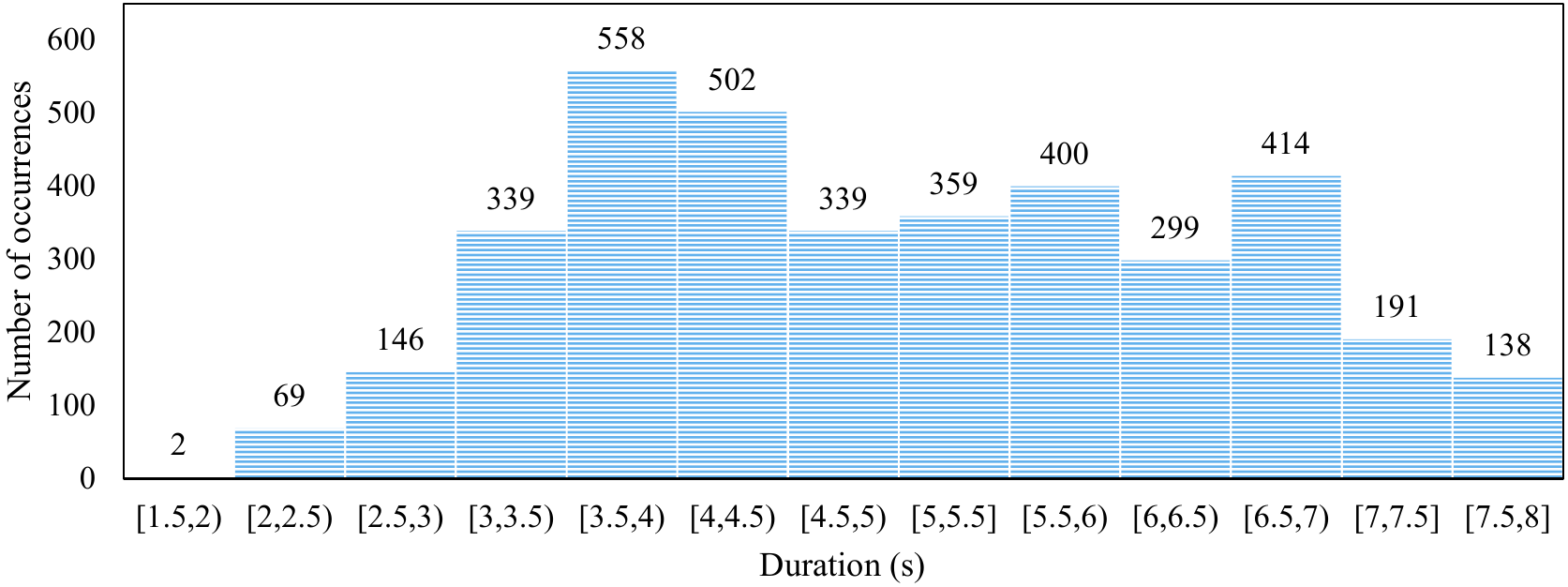}
    \caption{The statistical distribution of utterance duration. The shortest and longest duration is 1.8s and 8.0s, respectively.}
    \label{fig:duration}
\end{figure}

\begin{figure}[!h]
    \centering
    \setlength{\abovecaptionskip}{0.1cm}
    \setlength{\belowcaptionskip}{-0.5cm}
    \includegraphics[width=\linewidth]{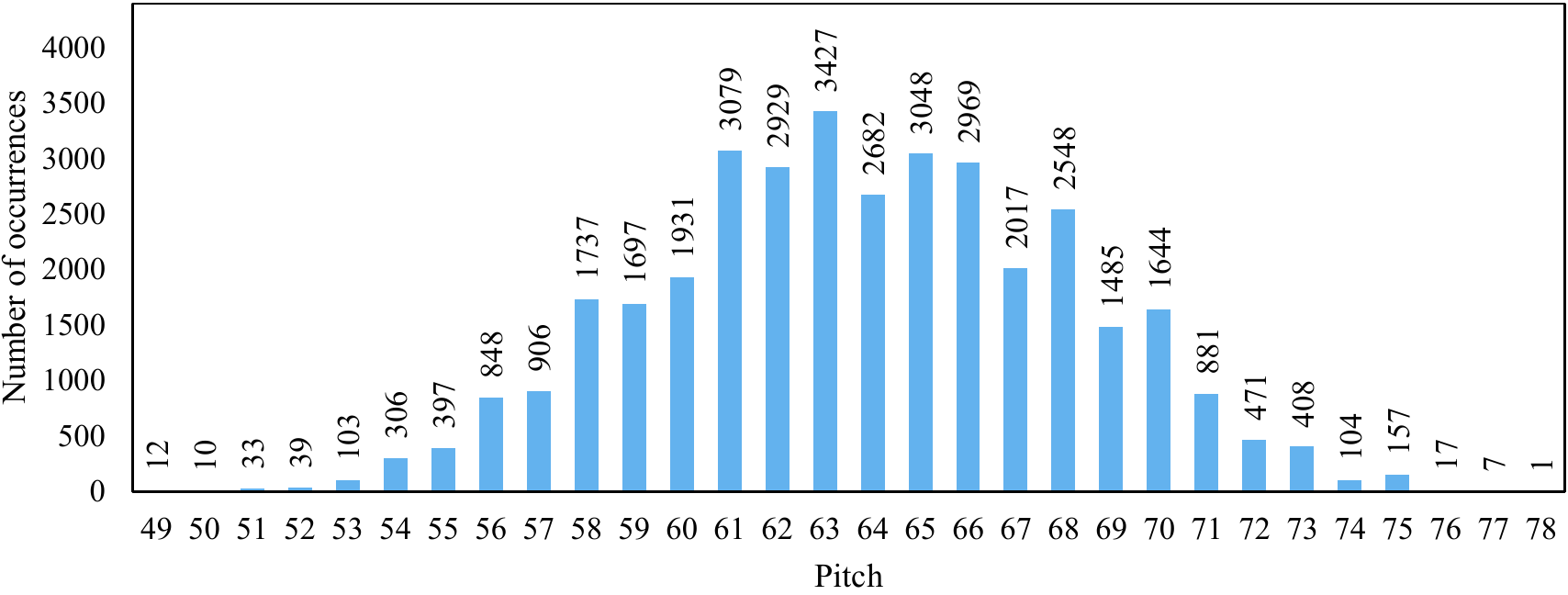}
    \caption{The statistical distribution of pitch. Pitch is presented as MIDI note number, where A4=69=440 Hz.}
    \label{fig:note}
\end{figure}

\begin{figure*}[t]
    \centering
    \setlength{\abovecaptionskip}{0.cm}
    \setlength{\belowcaptionskip}{-0.4cm}
    \includegraphics[width=\linewidth]{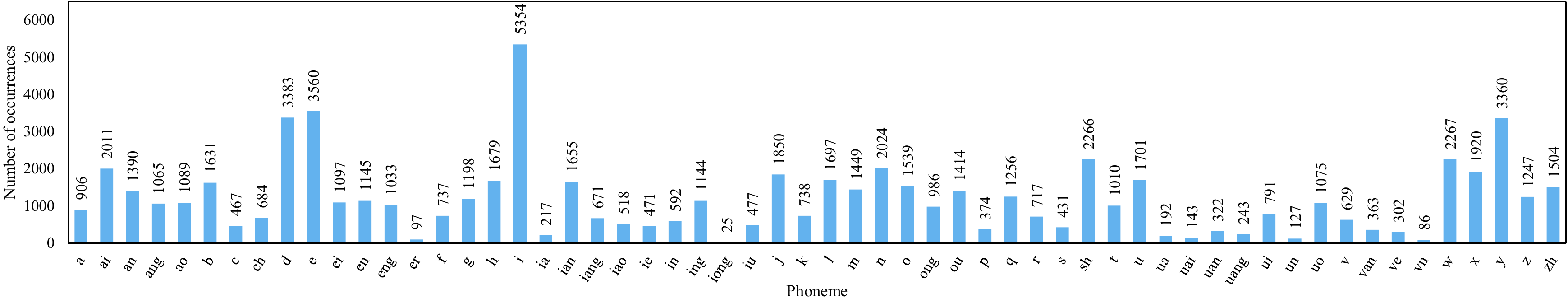}
    \caption{The statistical distribution of phonemes.}
    \label{fig:phoneme}
\end{figure*}

\subsection{Statistics of Opencpop}
\label{sc:statistics}

As introduced in Section~\ref{sc:song_selection}, to make the Opencpop cover a wide range of BPM, a song with the infrequent BPM would take the priority to be added to the song list of Opencpop. The final BPM distribution is shown in Fig.~\ref{fig:bpm}. As can be seen, Opencpop contains songs with various BPM that ranges from 40 to 130. Most songs' BPM falls within the range of 58 to 94, which is a popular BPM range of Mandarin Pop music. This wide 
coverage of BPM allows a system trained with Opencpop to deal with various BPM conditions.

The duration distribution of segmented utterances is shown in Fig.~\ref{fig:duration}. The duration of most utterances ranges from 3 seconds to 7 seconds. The note pitch distribution and phoneme distribution are presented in Fig.~\ref{fig:note} and Fig.~\ref{fig:phoneme}, respectively. The notes are mainly distributed between MIDI note number 56 (G\#3, 196~Hz) to 70 (B4, 494~Hz). As for the phoneme distribution, Opencpop covers all Mandarin phonemes, with the lowest frequency of phoneme ``iong“ which occurs 25 times in total, which can qualify an SVS system to handle all phonemes. 

\vspace{-0.3cm}
\section{Baseline systems and experiments}
\label{sc:experiments}
To verify the quality of Opencpop in the SVS task, and also to give a performance baseline in this task, several experiments towards the SVS task on the Opencpop database are conducted in this section.

\subsection{Methods}

A typical two-stage singing voice synthesis framework, which consists of an acoustic model and a vocoder is adopted in the experiments. In practice, Fastspeech2~\cite{ren2020fastspeech2} and HiFi-GAN~\cite{kong2020hifi}, which are two popular models for spectrogram synthesis and waveform reconstruction respectively, are utilized in this work.

Fastspeech2, which is a fully-feedforward architecture based on Transformer blocks~\cite{li2019neural}, is taken as the baseline in this paper. Different from the Fastspeech2-based TTS task, here, the prediction of energy and pitch is not considered. Instead, the input of the encoder is the concatenation of phoneme embedding and note embedding. As for the duration prediction, in addition to the phoneme level loss, an extra note level duration loss is added to provide the note-level duration constraint during the training process. The input of duration predictors is not only the phoneme embedding as in Fastspeech2 but also concatenates with other information, including note duration, slur indicator, phoneme position, and the phoneme number within a note.


While the Transformer block is good at modeling long-range global context, it has limited capability in capturing fine-grained local context. To face this limitation, Conformer thus came into being and has shown promising performance in the TTS task~\cite{guo2021recent}. In this work, we replace the Transformer blocks in Fastspeech2 with Conformer blocks, leading to a Conformer-based SVS model so-called \textit{CpopSing}. 

The Mel-prediction in Fastspeech2 is trained with L1 loss. However, it is well known that acoustic models trained with L1 or L2 loss suffer from over-smoothness of synthesized acoustic features, which could lead to singing voices with poor quality. Therefore, in addition to L1 loss, an adversarial training method is used during the training of CpopSing. This adversarial training method is similar to the sub-frequency adversarial loss in HifiSinger~\cite{chen2020hifisinger} but with an extra multi-length adversarial loss on the spectrogram.

In addition to CpopSing, with only L1 loss as the Mel reconstruction loss, a Transformed-based model and a Conformed-based model are also compared in the experiments. For convenience, these two models are referred to as \textit{TFSing} and \textit{CFSing} respectively, hereafter. 

\vspace{-0.2cm}
\subsection{Implementation details}
As a baseline for the Opencpop-based SVS task, here, we did not take the raw sampling rate 44,100 Hz as the target sampling rate. Instead, following the recent SVS research, the audio was down-sampled to 22,050 Hz, and then was represented by 80-band  Mel-scale spectrograms with the frame-shift of 12.5 ms. During the waveform reconstruction, the vocoder was trained with the real audio-spectrogram pairs, and then was fine-tuned with synthesized spectrograms paired with real audio.

\vspace{-0.2cm}
\subsection{Results}
Following~\cite{xue2021learn2sing}, evaluation metrics, i.e., F0 Root Mean Square Error (F0-RMSE), F0 Pearson Correlation Coefficient (F0-PCC), and duration accuracy (duracc) are used to evaluate the synthesized results objectively. To match the length difference between the ground-truth singing voice and the generated voice, the calculation of F0-RMSE and F0-PCC is conducted on generated singing voices that were created based on the ground-truth phoneme duration. Besides, to verify the quality of the synthesized singing voices subjectively, a Mean Opinion Score (MOS) test is also performed. In practice, all synthesized samples of the test set are used in the objective evaluation, while 30 samples are randomly chosen from the test set in the MOS test. The score of MOS test ranges from $1$ to $5$, in which $1$ means very bad and $5$ means excellent. Each audio is rated by 20 listeners.

The results are shown in Table~\ref{tb:results}. As shown, training on Opencpop, CpopSing achieves 3.70 MOS, falling in the range that near to \textit{good}, indicating the reliability of the proposed database Opencpop. On all evaluation metrics, except for duracc, CpopSing and CFSing achieve better performance than TFSing. In terms of objective metrics, CpopSing and CFSing show similar performance, while CpopSing outperforms CFSing on the subjective evaluation. These results indicate the effectiveness of Conformer blocks and Mel-based adversarial training method on the singing voice synthesis. From the generated results, baseline models showed limited capability in handling the pitch that falls in a long tail distribution in terms of pitch frequency in the training set\footnote{Please refer to the demo page for audio examples with spectrum: \url{https://wenet.org.cn/opencpop/resources/testset/}}. How to deal with this issue raised by the long tail distribution and even to perform a song with note pitch that is beyond the training set pitch distribution should be considered in future work.

\begin{table}
\centering
\footnotesize
\setlength{\abovecaptionskip}{0.1cm}%
\caption{Objective and subjective performance of different models on Opencpop. Lower F0-RMSE value means better performance. As for other metrics, higher value means better. ``gt dur" means the ground-truth phoneme duration is used in the synthesizing, while the synthesizing in ``predict dur" is based on the phoneme duration predictor.}
\setlength{\tabcolsep}{0.9mm}
\begin{tabular}{lccccc} 
\toprule
\multirow{2}{*}{Method} & \multirow{2}{*}{F0-RMSE} & \multirow{2}{*}{F0-PCC} & \multirow{2}{*}{duracc} & \multicolumn{2}{c}{MOS} \\ \cmidrule{5-6}
  &        &       &    &  predict dur  &   gt dur   \\  \midrule
TFSing & 29.99 & 0.893 & 0.885 & 3.45$\pm$0.05 & 3.51$\pm$0.05  \\
CFSing & 26.72 & 0.903 & 0.884 & 3.67$\pm$0.04 &  3.73$\pm$0.05 \\
CpopSing & 27.58 & 0.904 & 0.879 & 3.70$\pm$0.05 & 3.76$\pm$0.05  \\ \midrule
Ground Truth & --- & --- & --- &  \multicolumn{2}{c}{4.51$\pm$0.03}  \\
\bottomrule
\label{tb:results}
\vspace{-0.8cm}
\end{tabular}
\end{table}

\vspace{-0.2cm}
\section{Conclusions}
\label{sc:conclusions}
In this paper, we introduced the Opencpop corpus designed primarily for SVS systems. All recordings are phonetically labeled with phoneme boundaries and note boundaries manually. To our knowledge, this is the first open accessible high-quality Mandarin singing corpus with manual annotation. Opencpop can open up a number of possibilities for further research in the areas of SVS. 

\vspace{-0.2cm}
\section{Acknowledgements}
The authors thank contributors of the WeNet Open Source Community for maintaining this awesome community, especially to the primary maintainer Binbin Zhang for helping the release of Opencpop on the WeNet community. The authors also thank Mingming Fu, Tianyao Bai, Chen Li, and their colleagues in the audio department of NetEase Leihuo for their guidance and support in music basics.

\bibliographystyle{IEEEtran}

\bibliography{main}

\begin{thebibliography}{10}
\providecommand{\url}[1]{#1}
\csname url@samestyle\endcsname
\providecommand{\newblock}{\relax}
\providecommand{\bibinfo}[2]{#2}
\providecommand{\BIBentrySTDinterwordspacing}{\spaceskip=0pt\relax}
\providecommand{\BIBentryALTinterwordstretchfactor}{4}
\providecommand{\BIBentryALTinterwordspacing}{\spaceskip=\fontdimen2\font plus
\BIBentryALTinterwordstretchfactor\fontdimen3\font minus
  \fontdimen4\font\relax}
\providecommand{\BIBforeignlanguage}[2]{{%
\expandafter\ifx\csname l@#1\endcsname\relax
\typeout{** WARNING: IEEEtran.bst: No hyphenation pattern has been}%
\typeout{** loaded for the language `#1'. Using the pattern for}%
\typeout{** the default language instead.}%
\else
\language=\csname l@#1\endcsname
\fi
#2}}
\providecommand{\BIBdecl}{\relax}
\BIBdecl

\bibitem{wang2017tacotron}
Y.~Wang, R.~Skerry-Ryan, D.~Stanton, Y.~Wu, R.~J. Weiss, N.~Jaitly, Z.~Yang
  \emph{et~al.}, ``{Tacotron: Towards end-to-end speech synthesis},'' in
  \emph{Proc. INTERSPEECH}, 2017, pp. 4006--4010.

\bibitem{shen2017natural}
J.~Shen, R.~Pang, R.~J. Weiss, M.~Schuster, N.~Jaitly, Z.~Yang, Z.~Chen,
  Y.~Zhang, Y.~Wang, R.~Skerry-Ryan \emph{et~al.}, ``{Natural TTS synthesis by
  conditioning wavenet on mel spectrogram predictions},'' \emph{arXiv preprint
  arXiv:1712.05884}, 2017.

\bibitem{li2019neural}
N.~Li, S.~Liu, Y.~Liu, S.~Zhao, and M.~Liu, ``Neural speech synthesis with
  transformer network,'' in \emph{Proceedings of the AAAI Conference on
  Artificial Intelligence}, vol.~33, no.~01, 2019, pp. 6706--6713.

\bibitem{sotelo2017char2wav}
J.~Sotelo, S.~Mehri, K.~Kumar, J.~F. Santos, K.~Kastner, A.~Courville, and
  Y.~Bengio, ``{Char2wav: End-to-end speech synthesis},'' in \emph{Proc. ICLR
  workshop}, 2017.

\bibitem{ren2020fastspeech2}
Y.~Ren, C.~Hu, X.~Tan, T.~Qin, S.~Zhao, Z.~Zhao, and T.-Y. Liu, ``Fastspeech 2:
  Fast and high-quality end-to-end text to speech,'' \emph{arXiv preprint
  arXiv:2006.04558}, 2020.

\bibitem{yu2019durian}
C.~Yu, H.~Lu, N.~Hu, M.~Yu, C.~Weng, K.~Xu, P.~Liu, D.~Tuo, S.~Kang, G.~Lei
  \emph{et~al.}, ``Durian: Duration informed attention network for multimodal
  synthesis,'' \emph{arXiv preprint arXiv:1909.01700}, 2019.

\bibitem{miao2021efficienttts}
C.~Miao, L.~Shuang, Z.~Liu, C.~Minchuan, J.~Ma, S.~Wang, and J.~Xiao,
  ``Efficienttts: An efficient and high-quality text-to-speech architecture,''
  in \emph{International Conference on Machine Learning}.\hskip 1em plus 0.5em
  minus 0.4em\relax PMLR, 2021, pp. 7700--7709.

\bibitem{oord2016wavenet}
A.~v.~d. Oord, S.~Dieleman, H.~Zen, K.~Simonyan, O.~Vinyals, A.~Graves,
  N.~Kalchbrenner, A.~Senior, and K.~Kavukcuoglu, ``Wavenet: A generative model
  for raw audio,'' \emph{arXiv preprint arXiv:1609.03499}, 2016.

\bibitem{kim2018flowavenet}
S.~Kim, S.-g. Lee, J.~Song, J.~Kim, and S.~Yoon, ``Flowavenet: A generative
  flow for raw audio,'' \emph{arXiv preprint arXiv:1811.02155}, 2018.

\bibitem{prenger2019waveglow}
R.~Prenger, R.~Valle, and B.~Catanzaro, ``Waveglow: A flow-based generative
  network for speech synthesis,'' in \emph{ICASSP 2019-2019 IEEE International
  Conference on Acoustics, Speech and Signal Processing (ICASSP)}.\hskip 1em
  plus 0.5em minus 0.4em\relax IEEE, 2019, pp. 3617--3621.

\bibitem{kong2020hifi}
J.~Kong, J.~Kim, and J.~Bae, ``Hifi-gan: Generative adversarial networks for
  efficient and high fidelity speech synthesis,'' \emph{arXiv preprint
  arXiv:2010.05646}, 2020.

\bibitem{donahue2020end}
J.~Donahue, S.~Dieleman, M.~Binkowski, E.~Elsen, and K.~Simonyan, ``End-to-end
  adversarial text-to-speech,'' in \emph{International Conference on Learning
  Representations}, 2020.

\bibitem{weiss2021wave}
R.~J. Weiss, R.~Skerry-Ryan, E.~Battenberg, S.~Mariooryad, and D.~P. Kingma,
  ``Wave-tacotron: Spectrogram-free end-to-end text-to-speech synthesis,'' in
  \emph{ICASSP 2021-2021 IEEE International Conference on Acoustics, Speech and
  Signal Processing (ICASSP)}.\hskip 1em plus 0.5em minus 0.4em\relax IEEE,
  2021, pp. 5679--5683.

\bibitem{kim2021conditional}
J.~Kim, J.~Kong, and J.~Son, ``Conditional variational autoencoder with
  adversarial learning for end-to-end text-to-speech,'' \emph{arXiv preprint
  arXiv:2106.06103}, 2021.

\bibitem{skerry2018towards}
R.~Skerry-Ryan, E.~Battenberg, Y.~Xiao, Y.~Wang, D.~Stanton, J.~Shor, R.~Weiss,
  R.~Clark, and R.~A. Saurous, ``Towards end-to-end prosody transfer for
  expressive speech synthesis with tacotron,'' in \emph{international
  conference on machine learning}.\hskip 1em plus 0.5em minus 0.4em\relax PMLR,
  2018, pp. 4693--4702.

\bibitem{wang2018style}
Y.~Wang, D.~Stanton, Y.~Zhang, R.~Skerry-Ryan, E.~Battenberg, J.~Shor, Y.~Xiao,
  F.~Ren, Y.~Jia, and R.~A. Saurous, ``Style tokens: Unsupervised style
  modeling, control and transfer in end-to-end speech synthesis,'' \emph{arXiv
  preprint arXiv:1803.09017}, 2018.

\bibitem{lee2017emotional}
Y.~Lee, A.~Rabiee, and S.-Y. Lee, ``Emotional end-to-end neural speech
  synthesizer,'' \emph{arXiv preprint arXiv:1711.05447}, 2017.

\bibitem{rabiee2019adjusting}
A.~Rabiee, T.-H. Kim, and S.-Y. Lee, ``Adjusting pleasure-arousal-dominance for
  continuous emotional text-to-speech synthesizer,'' \emph{arXiv preprint
  arXiv:1906.05507}, 2019.

\bibitem{zhu2019controlling}
X.~Zhu, S.~Yang, G.~Yang, and L.~Xie, ``Controlling emotion strength with
  relative attribute for end-to-end speech synthesis,'' in \emph{2019 IEEE
  Automatic Speech Recognition and Understanding Workshop (ASRU)}.\hskip 1em
  plus 0.5em minus 0.4em\relax IEEE, 2019, pp. 192--199.

\bibitem{cai2020emotion}
X.~Cai, D.~Dai, Z.~Wu, X.~Li, J.~Li, and H.~Meng, ``Emotion controllable speech
  synthesis using emotion-unlabeled dataset with the assistance of cross-domain
  speech emotion recognition,'' \emph{arXiv preprint arXiv:2010.13350}, 2020.

\bibitem{um2020emotional}
S.-Y. Um, S.~Oh, K.~Byun, I.~Jang, C.~Ahn, and H.-G. Kang, ``Emotional speech
  synthesis with rich and granularized control,'' in \emph{ICASSP 2020-2020
  IEEE International Conference on Acoustics, Speech and Signal Processing
  (ICASSP)}.\hskip 1em plus 0.5em minus 0.4em\relax IEEE, 2020, pp. 7254--7258.

\bibitem{li2021controllable}
T.~Li, X.~Wang, Q.~Xie, Z.~Wang, and L.~Xie, ``Controllable cross-speaker
  emotion transfer for end-to-end speech synthesis,'' \emph{arXiv preprint
  arXiv:2109.06733}, 2021.

\bibitem{chen2020hifisinger}
J.~Chen, X.~Tan, J.~Luan, T.~Qin, and T.-Y. Liu, ``Hifisinger: Towards
  high-fidelity neural singing voice synthesis,'' \emph{arXiv preprint
  arXiv:2009.01776}, 2020.

\bibitem{liu2021efficientsing}
Z.~Liu, C.~Miao, Q.~Zhu, M.~Chen, J.~Ma, S.~Wang, and J.~Xiao, ``Efficientsing:
  A chinese singing voice synthesis system using duration-free acoustic model
  and hifi-gan vocoder,'' \emph{Proc. Interspeech 2021}, pp. 1609--1613, 2021.

\bibitem{ljspeech17}
K.~Ito and L.~Johnson, ``The lj speech dataset,''
  \url{https://keithito.com/LJ-Speech-Dataset/}, 2017.

\bibitem{veaux2016superseded}
C.~Veaux, J.~Yamagishi, K.~MacDonald \emph{et~al.}, ``Superseded-cstr vctk
  corpus: English multi-speaker corpus for cstr voice cloning toolkit,'' 2016.

\bibitem{shi2020aishell}
Y.~Shi, H.~Bu, X.~Xu, S.~Zhang, and M.~Li, ``Aishell-3: A multi-speaker
  mandarin tts corpus and the baselines,'' \emph{arXiv preprint
  arXiv:2010.11567}, 2020.

\bibitem{guo2021didispeech}
T.~Guo, C.~Wen, D.~Jiang, N.~Luo, R.~Zhang, S.~Zhao, W.~Li, C.~Gong, W.~Zou,
  K.~Han \emph{et~al.}, ``Didispeech: A large scale mandarin speech corpus,''
  in \emph{ICASSP 2021-2021 IEEE International Conference on Acoustics, Speech
  and Signal Processing (ICASSP)}.\hskip 1em plus 0.5em minus 0.4em\relax IEEE,
  2021, pp. 6968--6972.

\bibitem{duan2013nus}
Z.~Duan, H.~Fang, B.~Li, K.~C. Sim, and Y.~Wang, ``The nus sung and spoken
  lyrics corpus: A quantitative comparison of singing and speech,'' in
  \emph{2013 Asia-Pacific Signal and Information Processing Association Annual
  Summit and Conference}.\hskip 1em plus 0.5em minus 0.4em\relax IEEE, 2013,
  pp. 1--9.

\bibitem{sharma2021nhss}
B.~Sharma, X.~Gao, K.~Vijayan, X.~Tian, and H.~Li, ``Nhss: A speech and singing
  parallel database,'' \emph{Speech Communication}, vol. 133, pp. 9--22, 2021.

\bibitem{tamaru2020jvs}
H.~Tamaru, S.~Takamichi, N.~Tanji, and H.~Saruwatari, ``Jvs-music: Japanese
  multispeaker singing-voice corpus,'' \emph{arXiv preprint arXiv:2001.07044},
  2020.

\bibitem{lu2020xiaoicesing}
P.~Lu, J.~Wu, J.~Luan, X.~Tan, and L.~Zhou, ``Xiaoicesing: A high-quality and
  integrated singing voice synthesis system,'' \emph{arXiv preprint
  arXiv:2006.06261}, 2020.

\bibitem{huang2021multi}
R.~Huang, F.~Chen, Y.~Ren, J.~Liu, C.~Cui, and Z.~Zhao, ``Multi-singer: Fast
  multi-singer singing voice vocoder with a large-scale corpus,'' in
  \emph{Proceedings of the 29th ACM International Conference on Multimedia},
  2021, pp. 3945--3954.

\bibitem{liu2021diffsinger}
J.~Liu, C.~Li, Y.~Ren, F.~Chen, P.~Liu, and Z.~Zhao, ``Diffsinger: Singing
  voice synthesis via shallow diffusion mechanism,'' \emph{arXiv preprint
  arXiv:2105.02446}, vol.~2, 2021.

\bibitem{boersma2001praat}
P.~Boersma, ``Praat, a system for doing phonetics by computer,'' \emph{Glot.
  Int.}, vol.~5, no.~9, pp. 341--345, 2001.

\bibitem{guo2021recent}
P.~Guo, F.~Boyer, X.~Chang, T.~Hayashi, Y.~Higuchi, H.~Inaguma, N.~Kamo, C.~Li,
  D.~Garcia-Romero, J.~Shi \emph{et~al.}, ``Recent developments on espnet
  toolkit boosted by conformer,'' in \emph{ICASSP 2021-2021 IEEE International
  Conference on Acoustics, Speech and Signal Processing (ICASSP)}.\hskip 1em
  plus 0.5em minus 0.4em\relax IEEE, 2021, pp. 5874--5878.

\bibitem{xue2021learn2sing}
H.~Xue, S.~Yang, Y.~Lei, L.~Xie, and X.~Li, ``Learn2sing: Target speaker
  singing voice synthesis by learning from a singing teacher,'' in \emph{2021
  IEEE Spoken Language Technology Workshop (SLT)}.\hskip 1em plus 0.5em minus
  0.4em\relax IEEE, 2021, pp. 522--529.

\end{thebibliography}

\end{document}